\documentclass[aps,12pt, a4paper,preprint]{revtex4-1}

\usepackage[normalem]{ulem}
\usepackage[utf8]{inputenc}
\usepackage[T1]{fontenc}
\usepackage{amsmath}
\usepackage{amsfonts}
\usepackage{amssymb}
\usepackage{graphics,graphicx}

\usepackage{graphicx,color}
\usepackage{subfigure} 
\usepackage{wrapfig} 
\usepackage{epstopdf}
\graphicspath{{figuras/}}

\usepackage[breaklinks=true]{hyperref}
\usepackage{setspace}

\newcommand{\bes}{\begin{subequations}}
\newcommand{\ees}{\end{subequations}}
\def\ben{\begin{eqnarray}}
\def\een{\end{eqnarray}}
\def\be{\begin{equation}}
\def\ee{\end{equation}}

\begin{document}

\title{Shear viscosity {from black holes in} generalized scalar-tensor theories {in}  arbitrary dimensions}

\author{Mois\'es Bravo-Gaete}
\email{mbravo-at-ucm.cl} \affiliation{Facultad de Ciencias
B\'asicas, Universidad Cat\'olica del Maule, Casilla 617, Talca,
Chile.}

\author{Fabiano F. Santos}
\email{fabiano.ffs23-at-gmail.com}
\author{Henrique Boschi-Filho}
\email{boschi-at-if.ufrj.br}  
 \affiliation{Instituto de F\'{\i}sica, Universidade Federal do Rio de Janeiro, 21.941-972, Rio de Janeiro - RJ, Brazil.}

\begin{abstract}
In higher dimensions, we study  Degenerate-Higher-Order-Scalar-Tensor theories and we derive solutions that resemble the  Schwarzschild Anti-de Sitter black holes. We compute their thermodynamic quantities following the Wald formalism, satisfying the First Law of Thermodynamics and a higher dimensional Smarr relation. Constructing a Noether charge with a suitable choice of a space-like Killing vector, we obtain the shear viscosity of the non-gravitational dual field theory, where for a suitable choice of the couplings functions, the Kovtun-Son-Starinets bound is violated.  These results are  corroborated by the calculation of the Green's functions following the Kubo formalism. 
\end{abstract}

\maketitle
\newpage

\section{Introduction}
\label{intro}
General relativity (GR) provides the standard description of gravity. Extensions of GR have been  considered in the literature as gravitational alternatives, for instance, for unified descriptions of inflation and  cosmic acceleration. Various proposals include $F(R)$  gravity, scalar-tensor theories, string-inspired and Gauss-Bonnet theories \cite{Afonso:2007zz,Moraes:2016gpe,Heisenberg:2018vsk,Gasperini:2007zz,Sotiriou:2008rp,Karakasis:2021lnq,Karakasis:2021rpn}. Some of these models might be consistent with local tests, and the occurrence of finite-time future singularities in modified gravity may be cured by the addition of higher-derivative  terms. For a review, see {\it e. g.} \cite{Nojiri:2010wj}. 
In particular, at the seventies, Horndeski constructed a four-dimensional scalar-tensor theory wherein the equations of motions are at most of the second order in the derivatives of the field functions  \cite{Horndeski:1974wa}.
Motivated with the above, in recent years a new class of scalar-tensor theories of gravity that extend Horndeski, or “generalized Galileon,” models have been proposed. Despite possessing equations of motion of higher-order derivatives, the propagating degrees of freedom satisfy second-order equations of motion and are thus free from Ostrogradski instabilities  \cite{Gleyzes:2014dya}.

Astrophysical implications \cite{Riess:1998cb,Perlmutter:1998np,Abbott:2016blz}  have yielded further motivation to study theories of gravity beyond Horndeski proposal, channeling in a model denominated as Degenerate-Higher-Order-Scalar-Tensor (DHOST) theory, also avoiding Ostrogradsky instability due to its degeneracy property \cite{BenAchour:2016fzp,Motohashi:2016ftl}. 
It is important to note that these degenerate theories have allowed the exploration of four- dimensional regular black holes \cite{Babichev:2020qpr,Baake:2021jzv}, rotating black holes stealth \cite{Baake:2021kyg} and three-dimensional spinning configurations \cite{Baake:2020tgk}. As far as we know, the extension of these solutions for higher-dimensional theories is still an open problem, which we address in this work.

{On the other hand, the Anti-de Sitter/Conformal Field Theory} (AdS/CFT)  correspondence  \cite{Maldacena:1997re,Gubser:1998bc,Witten:1998qj} is a relation between a gravitational theory in $D$- spacetime dimensions and a field theory in flat $(D-1)$- dimensions (without gravity). In its most general form, it is known as a gauge/gravity duality. One nice property of this duality is that when the field theory is strongly coupled the gravitational dual is weakly coupled and vice-versa. This property opens a large window of applications in many different areas. In particular, it gives support to study the dynamics of tightly coupled  systems, especially the transport coefficients  from condensed matter and hydrodynamics to the quark-gluon plasma formed at  relativistic heavy-ion  collisions \cite{Jiang:2017imk,Cisterna:2017jmv,Baggioli:2017ojd,Hartnoll:2009sz,Kovtun:2003wp,Kovtun:2004de,Sachdev:2011wg,Sadeghi:2018vrf,Sil:2016jmc}. One of these coefficients is the well-known shear viscosity $\eta$ \cite{Kovtun:2003wp,Kovtun:2004de}, calculated from holographic bottom-up models. Under this scenario, it is possible to compute the ratio between $\eta$ and the entropy density ${s}$, arising a conjecture about a universal bound, known as the Kovtun-Son-Starinets (KSS) bound, which reads
\cite{Kovtun:2003wp,Kovtun:2004de,Policastro:2001yc,Son:2002sd} 
\begin{equation}\label{KSS}
\frac{\eta}{s}\geq \frac{1}{4 \pi}, %
\end{equation}
being support in a variety of gravity dual models \cite{Buchel:2003tz,Buchel:2004qq,Benincasa:2006fu,Landsteiner:2007bd}. 

On the contrary, for some  examples which include unconventional superconducting systems \cite{Baggioli:2017ojd,Kovtun:2003wp}, the Einstein-Hilbert Gauss-Bonnet action in five dimensions \cite{Kats:2007mq,Brigante:2007nu}, the Horndeski Theory  \cite{Feng:2015oea, 
Brito:2019ose,Liu:2018hzo, Figueroa:2020tya,Bravo-Gaete:2021hlc}, as well as the DHOST theories in four dimensions \cite{Santos:2020lmb} the KSS bound is violated, at least for some choices of the relevant parameters of these models.

In this work, we consider DHOST theories in higher dimensions {($D\ge 3$)}. We construct hairy black hole solutions with planar geometry under this scenario and study their thermodynamic properties. Together with the above, in order to obtain the shear viscosity of the dual gauge theories living in lower-dimensional flat spaces, we employ two different methods for $D>3$. 
The first one is performed via the Wald formalism, with the construction of a Noether charge and an election for the space-like Killing vector \cite{Fan:2018qnt}. For the second,  we use the more traditional methods present in \cite{Son:2002sd} and \cite{Brigante:2007nu},  with the calculation of Green's functions and the use of the Kubo formula. The two formalisms generate the same expression for the shear viscosity of the dual gauge field theories, allowing violation of the KSS bound.

This paper is organized as follows: In Section \ref{v2},  we consider DHOST theories in higher dimensions {($D \geq 3$)} and obtain hairy black hole solutions with planar base manifolds in these spacetimes. In Section \ref{sec.termo}, through the Wald formalism \cite{Wald:1993nt,Iyer:1994ys} the thermodynamics of that solutions are explored, and in Section \ref{viscosity} we obtain the viscosity/entropy density ratio of the corresponding dual field theories for $D>3$,  showing that the KSS bound could be violated in these theories. {Some details of the equations of motion are presented in Appendix \ref{Appendix}. Further, in Appendix \ref{Kubo}, we reobtain the viscosity/ entropy density ratio using Green's functions and the Kubo formula, corroborating the results of Section \ref{viscosity}.  Finally, Section \ref{conclusions} is devoted to our conclusions and discussions.}

\section{DHOST theories and hairy black hole solutions in higher dimensions}\label{v2} 

To our knowledge, DHOST theories in $D=3$ and $D=4$ spacetime dimensions have been studied in Refs. \cite{Heisenberg:2018vsk,Gleyzes:2014dya,Motohashi:2016ftl,BenAchour:2016fzp,Babichev:2020qpr,Baake:2021jzv,Baake:2021kyg,Baake:2020tgk}. Here we generalize these previous approaches to the $D$-dimensional case ($D\ge 3$) defining the action as
\begin{eqnarray}
S[g_{\mu \nu},\phi]&=&\int d^{D}x\sqrt{-g}\, \mathcal{L},\label{action}
\end{eqnarray}
where the Lagrangian $\mathcal L$ reads 
\begin{eqnarray}
\mathcal{L}&=&\lambda_0 Z(X)+[1+\lambda_1 G(X)]R+\sum_{i=2}^{5} \lambda_i A_{i}(X)  {\mathcal{L}}_{i},\label{lagrangian}
\end{eqnarray}
with $X:=\partial_{\mu}\phi\,\partial^{\mu}\phi$ being the kinetic term {of the scalar field $\phi$}, and 
\begin{eqnarray}
{\mathcal{L}}_{2}&:=&(\Box\phi)^2-\phi_{\mu\nu}\phi^{\mu\nu},\qquad  {\mathcal{L}}_{3}:=\Box\phi\,\phi^{\mu}\phi_{\mu\nu}\phi^{\nu},\label{L23}\\
{\mathcal{L}}_{4}&:=&\phi^{\mu}\phi_{\mu\nu}\phi^{\nu\rho}\phi_{\rho}
,\qquad {\mathcal{L}}_{5}:=\left(\phi^{\mu}\phi_{\mu\nu}\phi^{\nu}\right)^2. \label{L45}
\end{eqnarray}
Here  $R$ is the scalar curvature, $\lambda_m$, with $m \in \{0,1,2,3,4,5\}$,  are convenient parameters to control the couplings between the functions $Z(X)$,  $G(X)$ and $A_i(X)$, with $i\in \{2,3,4,5\}$,
while we have defined  $\phi_{\mu}:=\nabla_{\mu}\phi, \, {\rm and} \;  \phi_{\mu\nu}:=\nabla_{\mu}\nabla_{\nu}\phi$. {For later convenience, we also define the derivative with respect to $X$,   $F_X:=dF/dX$, so that, for instance,} $Z_{X}:=dZ /dX$, $G_{X}:=dG/dX$, and $A_{i X}:=dA_{i}/dX$.
The equations of motions with respect to the metric $g_{\mu \nu}$ and the scalar field $\phi$  are given by
\begin{eqnarray}
\mathcal{E}_{\mu\nu}&:=&{\cal{G}}^{Z}_{\mu\nu}
+{\cal{G}}^{G}_{\mu\nu}+\sum_{i=2}^{5}{\cal{G}}^{(i)}_{\mu\nu}=0,\label{eq:emotion1}\\
{\mathcal{E}_{\phi}}&=&\nabla_{\mu}
\mathcal{J}^{\mu}=\nabla_{\mu}\left[\frac{\delta \cal{L}}{\delta
(\phi_{\mu})}-\nabla_{\nu}\left(\frac{\delta \cal{L}}{\delta (
\phi_{\mu \nu})}\right)\right]=0,\label{eq:emotion2}\\
\nonumber
\end{eqnarray}
where the expressions given in $\mathcal{E}_{\mu\nu}$ and $\mathcal{J}^{\mu}$ are reported in the Appendix \ref{Appendix}.

For the generalized scalar-tensor configuration Eqs.  (\ref{action})-(\ref{L45}), we consider the following higher-dimensional metric {\sl ansatz}: 
\begin{eqnarray}
ds^{2}&=&-h(r)dt^{2}+\frac{dr^{2}}{f(r)}+r^{2} \sum_{i=1}^{D-2} d x_i^2,\label{7} \\
\phi(t,r)&=&\psi(r),
\end{eqnarray}
where only a radial dependence for the scalar field $\phi$ is required, given that we are working on a planar base manifold. 
In order to simplify our computations, we also suppose that the kinetic term $X$ is a constant. This hypothesis implies that
\begin{equation}
X=g^{rr} (\psi')^2,\label{eq:X}  
\end{equation}
and the square of the derivative of the scalar field $\phi$ can be cast as
\begin{equation}
(\psi')^{2}=\frac{X}{f},\label{11}
\end{equation}
where $(')$ denotes the derivative with respect to the radial coordinate $r$. {Following the steps performed in \cite{Baake:2020tgk,Santos:2020lmb},  {we fix the function $A_5$ as 
\begin{subequations}
\begin{eqnarray}
\lambda_5 A_5=\frac{\left(2 \lambda_2A_2+X \lambda_3 A_3+4\lambda_1G_X\right)^2}{2X(1+\lambda_1G+\lambda_2XA_2)}-\left(\frac{\lambda_3A_3+\lambda_4A_4}{X}\right),
\label{A5}
\end{eqnarray}
{or 
\begin{eqnarray}
\lambda_5 A_5=\frac{1}{X}\left( \frac{\left[\mathcal{Z}_2(X) \right]^2}{2 \mathcal{Z}_1(X)}
-\mathcal{Z}_3(X)\right),
\label{A5}
\end{eqnarray}}
{defining the functions} 
\label{defZ}
\begin{eqnarray}
\mathcal{Z}_1 (X) &=& 1+\lambda_1G(X) + X \lambda_2A_2(X), \label{defZ1}\\
\mathcal{Z}_2(X) &=& 2 \lambda_2A_2(X) + \lambda_3 X A_3(X) + 4 \lambda_1G_X(X),  \label{defZ2}
\\
{
\mathcal{Z}_3(X)} &=& 
{\lambda_3 A_3(X) + \lambda_4 A_4(X).} \label{defZ3}
\end{eqnarray}
\end{subequations}
Then, a solution in higher dimensions $D\geq 3$ reads
\begin{eqnarray}
f(r)&=&h(r)=\frac{\lambda_0Z r^{2}}{ (D-1) (D-2) \, \mathcal{Z}_1}-\frac{M}{r^{D-3}},\label{10}
\end{eqnarray}
 where $M$ is a positive integration constant, as long as the coupling functions are related in the following form}
\begin{eqnarray}\label{eq:mastereq}
2 (D-2) \left(Z \mathcal{Z}_{1}\right)_{X}=(D-1) \mathcal{Z}_{2} Z,
\end{eqnarray}
{and the scalar field from (\ref{11}) can be obtained as
\begin{eqnarray}\label{eq:phi}
\phi(t,r)=\psi(r)=\pm \left(\dfrac{2 l}{D-1}\right) \sqrt{X} \ln\left[r^{\frac{D-3}{2}} \left(\frac{r}{l}+\sqrt{\frac{r^2}{l^{2}}-\frac{M}{r^{D-3}}}\right)\right].
\end{eqnarray}
Many commentaries can be carried out with respect to the solution (\ref{10})-(\ref{eq:phi}). First, the metric function $h=f$ resembles the well-known Schwarzschild-AdS {black hole} in {$D$}-{spacetime} dimensions. Second, the equation (\ref{eq:mastereq}) represents the extension of the particular cases found previously in four \cite{Santos:2020lmb} and three dimensions \cite{Baake:2020tgk}, where the scalar field is well defined on the location of the event horizon $r_h=(Ml^2)^{\frac{1}{D-1}}$ and in order to have a real and non trivial expression for $\phi$, we need $X >0$ for $r \geq r_h$}. Finally, {in order} to have {an}  asymptotically AdS black hole configuration, we will define the AdS radius $l$ as
\begin{equation}\label{eq:L2}
l^2=\frac{ (D-1)(D-2)\mathcal{Z}_1}{\lambda_0 Z},
\end{equation}
{and impose the constraint}
$$\frac{ \mathcal{Z}_1}{\lambda_0 Z}>0,$$
 to have a real expression for $l$. 
Summarizing, with the DHOST theory Eqs. (\ref{action})-(\ref{L45}) together with a constant kinetic term $X$ and the coupling functions $Z,G$ and the $A_{i}$'s satisfying the relation (\ref{eq:mastereq}), we can obtain a higher dimensional hairy solution with a planar base manifold given in Eqs.  (\ref{11})-(\ref{10}). In the following section,  we will derive the thermodynamic quantities corresponding to this solution.

\section{Thermodynamics of the hairy solution from the Wald formalism} 
\label{sec.termo}

Given the hairy higher-dimensional black hole  solution, found in the previous section,  to compute extensive thermodynamic quantities (these are the mass $\mathcal{M}$ and the  entropy $\mathcal{S}_{W}$), we will consider the Wald formalism \cite{Wald:1993nt,Iyer:1994ys}. {We start through the variation of the action (\ref{action})-(\ref{L45}) with respect to all the dynamical fields, which is
\begin{eqnarray*}
\delta S&=&\sqrt{-g} \big[{\cal{E}}_{\mu \nu} \delta g^{\mu \nu} + {\cal{E}}_\phi \delta \phi + \nabla_{\mu} J^{\mu} (\delta g, \delta \phi), \big]
\end{eqnarray*}
where, as before, ${\cal{E}}_{\mu \nu}$ and ${\cal{E}}_\phi$ are the equations of motions with respect to the metric and the scalar field. The surface term $J^{\mu}$ reads 
\begin{eqnarray}
J^{\mu}&=&\sqrt{-g}\Big[2\left(P^{\mu (\alpha\beta)
\gamma}\nabla_{\gamma}\delta g_{\alpha\beta}-\delta g_{\alpha\beta} \nabla_{\gamma}P^{\mu(\alpha\beta)\gamma}\right)+ \mathcal{J}^{\mu} \delta \phi
+\frac{\delta \cal{L}}{\delta (\phi_{\mu \nu})} \delta (\phi_{\nu})\nonumber\\
&-&\frac{1}{2}\frac{\delta \cal{L}}{\delta (\phi_{\mu \sigma})}
\phi^{\rho} \,
\delta g_{\sigma \rho}-\frac{1}{2}\frac{\delta \cal{L}}{\delta (\phi_{\sigma \mu})}\phi^{\rho} \,\delta g_{\sigma \rho}+\frac{1}{2}\frac{\delta \cal{L}}{\delta ( \phi_{\sigma
\rho})}\phi^{\mu}\,\delta g_{\sigma \rho}\Big],\label{eq:surface}
\end{eqnarray}
with $\mathcal{J}^{\mu}$ reported in the Appendix \ref{Appendix}}. {Further, in our case:
\begin{equation}
P^{\mu \nu \sigma \rho}=\frac{\delta \mathcal{L}}{\delta R_{\mu \nu \sigma \rho}}=\frac{1}{2}\,(1+\lambda_1G(X))\, \left(g^{\mu \sigma} g^{\nu \rho}-g^{\mu \rho} g^{\nu \sigma}\right), \label{P}    
\end{equation}
while that
\begin{eqnarray}
\frac{\delta \mathcal{L}}{\delta \phi_{\mu}}&=&2 \lambda_0Z_{X} \phi^{\mu}+2 \lambda_1G_{X} R \phi^{\mu}+2 \lambda_2A_{2X} \phi^{\mu}\left[ (\Box \phi)^{2}-
\phi_{\lambda \rho}\phi^{\lambda \rho}\right]\nonumber \\
&+&2 \lambda_3 A_{3X} \phi^{\mu}  \, \Box \phi\,  \phi^{\lambda}
\phi_{\lambda \rho} \phi^{\rho}+2 \lambda_3 A_3  \,\Box \phi \,
\phi^{\mu}_{\,\,\,\lambda}\phi^{\lambda}\nonumber \\
&+&2 \lambda_4 A_{4X} \phi^{\mu} \phi^{\sigma} \phi_{\sigma \rho}
\phi^{\rho \lambda}
\phi_{\lambda}+\lambda_4A_4(X) \big[\phi^{\mu}_{\,\,\rho} \phi^{\rho \lambda} \phi_{\lambda}+\phi^{\sigma} \phi_{\sigma \rho} \phi^{\rho \mu} \big]\nonumber \\
&+&2 \lambda_5 A_{5X} \phi^{\mu} \big( \phi^{\sigma} \phi_{\sigma
\rho} \phi^{\rho}\big)^2 +2 \lambda_5 A_5(X)\big(
\phi^{\sigma} \phi_{\sigma \rho} \phi^{\rho}\big)\big(\phi^{\mu \sigma}\phi_{\sigma}\nonumber \\
&+&\phi^{\sigma \mu} \phi_{\sigma}\big), \label{VarL1}
\end{eqnarray}
{and}
\begin{eqnarray}
\frac{\delta \mathcal{L}}{\delta \phi_{\mu \nu}}&=& 2\lambda_2 A_2  \left( g^{\mu \nu}-\phi^{\mu \nu}\right)+\lambda_3 A_3  \big(g^{\mu \nu} \phi^{\lambda} \phi_{\lambda \rho} \phi^{\rho}+\Box \phi \,\phi^{\mu} \phi^{\nu}\big)+\lambda_4 A_4(X) \big(\phi^{\mu} \phi^{\nu \rho} \phi_{\rho}\nonumber \\
&+&\phi^{\sigma} \phi_{\,\,\,\sigma}^{\mu}
\phi^{\nu}\big)+2 \lambda_5 A_5(X)
\phi^{\sigma} \phi_{\sigma \rho} \phi^{\rho}  \phi^{\mu}
\phi^{\nu}.\label{VarL2}
\end{eqnarray}
}Defining a $1$-form $J_{(1)}=J_{\mu} dx^{\mu}$ and its corresponding Hodge dual ${\Theta}_{(D-1)}=(-1)^{D+1}* J_{(1)},$
{together with considering a variation induced by an infinitesimal diffeomorphism $\delta x^{\mu}=\xi^{\mu}$, and making use of the equations of motions (\ref{eq:emotion1})-(\ref{eq:emotion2}), we have that
$J_{(D-1)}={\Theta}_{(D-1)}-i_{\xi}(*\mathcal{L})=d(*J_{(2)}),$
where $i_{\xi}$ is a contraction of the vector $\xi^{\mu}$  with the first index of $*\mathcal{L}$, and in our notations the subindex "$(p)$" corresponds to the fact that we are working with $p-$forms}. The above allows the definition of a $(D-2)-$form {${Q}_{(D-2)}=*{J}_{(2)}$ such that ${J}_{(D-1)}=d Q_{(D-2)}$}, where 
$$Q_{(D-2)}=Q_{\alpha_1 \alpha_2 \cdots \alpha_{D-2}}=\epsilon_{\alpha_1 \alpha_2 \cdots \alpha_{D-2} \mu \nu} Q^{\mu \nu},$$
with
\begin{eqnarray}
Q^{\mu \nu}&=&2P^{\mu\nu\rho\sigma}\nabla_\rho \xi_\sigma -4\xi_\sigma
\nabla_\rho P^{\mu\nu\rho\sigma}+\frac{\delta \cal{L}}{ \delta \phi_{\mu \sigma}} \phi^{\nu} \xi_{\sigma}\nonumber \\
&-&\frac{\delta \cal{L}}{ \delta \phi_{\nu \sigma}} \phi^{\mu}
\xi_{\sigma}, \label{eq:noether}
\end{eqnarray}
and $P^{\mu\nu\rho\sigma}$ and ${\delta \cal{L}}/{ \delta \phi_{\mu \sigma}}$ were given previously in Eqs. (\ref{P}) and (\ref{VarL2}), respectively. {Concretely, for the action (\ref{action})-(\ref{L45}) and using the fact from (\ref{11}) that $\delta (\phi')= -\sqrt{X} \delta f / (2 f^{3/2})$ (here we note that  $\mathcal{J}^{\mu} \delta \phi$ from (\ref{eq:surface}) vanishes after making use of the equations of motion), we find that $i_{\xi} {\Theta}_{(D-1)}$ as well as $Q_{(D-2)}$ read  
\begin{eqnarray*}
i_{\xi} {\Theta}_{(D-1)}&=& r^{D-3} \big[
-(D-2) \mathcal{Z}_1 \delta f+2(D-2) \delta f (1+\lambda_1 G) +r (1+\lambda_1 G) \delta (f')\big] {\Omega_{D-2}}, \\
Q_{(D-2)}&=& r^{D-3} \big[r (1+\lambda_1 G) f'+2 (D-2) f (1+\lambda_1 G-\mathcal{Z}_1)\big] {\Omega_{D-2}},
\end{eqnarray*}
and the variation of $Q_{(D-2)}$ takes the form
$$\delta Q_{(D-2)} =r^{D-3} \big[r (1+\lambda_1 G) \delta (f')+2 (D-2) \delta f (1+\lambda_1 G-\mathcal{Z}_1)\big] {\Omega_{D-2}},$$
where ${\Omega_{D-2}} $  is the finite volume of the $(D-2)-$dimensional compact angular base manifold. Finally, taking $\xi^{\mu}$ as a time-like Killing vector that is null on the location of the event horizon, denoted as $r_h$, the variation of the Hamiltonian reads 
\begin{eqnarray*}
\delta \mathcal{H}&=&\delta \int_{\mathcal{C}} J_{(D-1)} -\int_{\mathcal{C}} d \left(i_{\xi} \Theta_{(D-1)}\right)=\delta \int_{\mathcal{C}} d \left(Q_{(D-2)}\right) -\int_{\mathcal{C}} d \left(i_{\xi} \Theta_{(D-1)}\right)\\
&=&\int_{\Sigma^{(D-2)}}\left(\delta {Q}_{(D-2)}-i_{\xi} {\Theta}_{(D-1)}\right),
\end{eqnarray*}
where $\mathcal{C}$ and $\Sigma^{(D-2)}$ are a Cauchy surface and its boundary respectively, which has {two} components, located at infinity ($ \mathcal{H}_{\infty}$) and at the horizon ($\mathcal{H}_{+}$).} According to {the Wald formalism \cite{Wald:1993nt,Iyer:1994ys}}, the first law of black holes thermodynamics
\begin{equation}
\delta\mathcal{M}= T \delta \mathcal{S}_{W},\label{eq:first-law}
\end{equation}
is a consequence of $\delta \mathcal{H}_{\infty}=\delta \mathcal{H}_{+}$, where $\mathcal{M}$ and $\mathcal{S}_{W}$ denote the mass as well as the entropy}, while that {the Hawking Temperature $T$ reads
\begin{eqnarray}
T=\frac{\kappa}{2 \pi}\Big{|}_{r=r_h}&=&\frac{1}{2\pi}\sqrt{-\frac{1}{2}\left(\nabla_{\mu} \xi_{\nu}\right)\left(\nabla^{\mu} \xi^{\nu}\right)} \bigg|_{r=r_h}  \nonumber \\
&=&\frac{1}{4 \pi} \frac{\lambda_0 Z r_h}{ (D-2) \mathcal{Z}_{1}}=\frac{(D-1) r_h}{4 \pi l^2} ,\label{1t1}
\end{eqnarray}
{where} $\kappa$ is the surface gravity,  $r_h$ is the location of the event horizon and the AdS radius $l$ was defined previously in Eq.  (\ref{eq:L2}).} {With all the above, we have that
$$\delta {Q}_{(D-2)}-i_{\xi} {\Theta}_{(D-1)}=- (D-2) r^{D-3} \mathcal{Z}_1 \delta f,$$
where
$$\delta f=-\frac{\delta M}{ r^{D-3}}.$$}
Computing the respective variation of the solution (\ref{10})-(\ref{11}), at the infinity we can write
\begin{eqnarray}
\delta \mathcal{H}_{\infty}&=& \delta \mathcal{M}=(D-2) \mathcal{Z}_1  {\Omega_{D-2}} 
\delta M 
\end{eqnarray}
{so that the mass $\mathcal{M}$ takes the form}  
\begin{eqnarray}
&&\mathcal{M} = (D-2) \mathcal{Z}_1 M \Omega_{D-2}=\frac{ (D-2) \mathcal{Z}_1 r_h^{D-1} {\Omega_{D-2}} }{l^2}.\label{mass}
\end{eqnarray} 
{Note that the positivity of the physical mass $\mathcal{M}$ implies that $\mathcal{Z}_1 > 0$. This thermodynamic condition  will be important, as we will see below, at the moment to study the shear viscosity $\eta$.} {At the horizon, where from the metric function (\ref{10}) $$\delta M=\frac{(D-1) r_h^{D-2} \delta r_h}{l^{2}},$$
where $l^{2}$ is the AdS radius (\ref{eq:L2}),
} 
we have
\begin{eqnarray}
\delta \mathcal{H}_{+}=T \delta \mathcal{S}_{W} =T \,\delta\left(4 \pi {\Omega_{D-2}}  \mathcal{Z}_1 r_h^{D-2}  \right),
\end{eqnarray}
from which the entropy $\mathcal{S}_{W}$ takes the form
\begin{eqnarray}
\mathcal{S}_{W}=4 \pi {\Omega_{D-2}}  \mathcal{Z}_1 r_h^{D-2}.\label{entropy}
\end{eqnarray}
{Note that the condition  $\mathcal{Z}_1 >0$ from the positivity of the mass $\mathcal{M}$ from eq. \eqref{mass} guarantees the positivity of the entropy $\mathcal{S}_{W}$.}
 It is worth pointing out that besides the fulfills of the first law (\ref{eq:first-law}), a higher-dimensional Smarr relation \cite{Smarr:1972kt}
\begin{equation}\label{eq:smarr}
\mathcal{M}= \left(\frac{D-2}{D-1}\right)\,T \mathcal{S}_{W},   
\end{equation}
holds. 

\section{The viscosity/entropy density ratio through the Wald formalism}
 \label{viscosity}

After obtaining the thermodynamic quantities from the hairy higher-dimensional black hole solution, in particular for the Wald entropy $\mathcal{S}_{W}$ from (\ref{entropy}) we can obtain the entropy density $s$ in our set up, given by
\begin{equation}
 s=\frac{\mathcal{S}_{W}}{{\Omega_{D-2}}}={4 \pi r_h^{D-2} {\mathcal{Z}_1}}.\label{s}   
\end{equation}
In order to calculate the shear viscosity $\eta$, according to the procedure performed in \cite{Fan:2018qnt}, we first perform a transverse and traceless perturbation on the metric (\ref{7}) for $D >3$ with $h=f$, which reads 
\begin{equation}\label{pert_metric}
ds^2=-f(r) dt^2+\frac{dr^2}{f(r)}+2 r^2 \Psi(t,r) dx_1 dx_2+ r^2 \sum_{i=1}^{D-2} d x_{i}^2,
\end{equation}
 with the ansatz 
$$\Psi(t,r) =\zeta t + h_{x_1 x_2}(r),$$
where $\zeta$ is a constant identified as the gradient of the fluid velocity along the $x_1$ direction. {This perturbation} yields the following $(x_1,x_2)$-component of the linearized Einstein equations:  
\begin{eqnarray}\label{linear}
\left[\mathcal{Z}_{1}(X)r^{D-2} f (h_{x_1 x_2})'\right]'=0, 
\end{eqnarray}
and by using a space-like Killing vector $\partial_{x_1}=\xi^{\mu}\partial_{\mu}$, the charge $\sqrt{-g} {Q}^{r x_2}$, constructed through $Q^{\mu \nu}$ from (\ref{eq:noether}), becomes an integration constant \cite{Fan:2018qnt}, which reads   
\begin{equation}
 \sqrt{-g} {Q}^{r x_2}= \mathcal{Z}_{1} (X) \,r^{D-2} f (h_{x_1 x_2})'. 
\end{equation}\\
Imposing the ingoing horizon boundary condition
$$h_{x_1 x_2} = \zeta  \sqrt{\frac{1+\lambda_1G}{\mathcal{Z}_1}}\,\frac{\log (r-r_h)}{4 \pi T}+\cdots,$$\\
as well as a Taylor expansion in the near horizon region $r_h$
$$h=f=4\pi T (r-r_h)+\cdots,$$
where $T$ is the Hawking temperature {given previously in (\ref{1t1})}, we have: 
$$ {\sqrt{-g} {Q}^{r x_2} = \zeta  \mathcal{Z}_{1}\, \sqrt{\frac{1+\lambda_1G}{{\mathcal{Z}_1}}}\, r_h^{D-2}=   \zeta \left(\frac{1}{4 \pi}  \sqrt{\frac{1+\lambda_1G}{{\mathcal{Z}_1}}} s\right),  }$$
{where $s$ was given in (\ref{s}). Following the steps from \cite{Fan:2018qnt}, the shear viscosity $\eta$ can be obtained in the following way}
\begin{equation}{\eta=\frac{\partial (\sqrt{-g} {Q}^{r x_2})}{\partial \zeta}=\frac{1}{4 \pi}  \sqrt{\frac{1+\lambda_1G}{{\mathcal{Z}_1}}} s.}
\end{equation}
{Since the shear viscosity $\eta$ is real and non-negative, and the fact that $\mathcal{Z}_1>0$, from the positivity of the mass $\mathcal{M}$ \eqref{mass} as well as the entropy $\mathcal{S}_{W}$ \eqref{entropy}, we conclude that $1+\lambda_1G \ge 0$.}
Then, the viscosity/entropy density ratio takes the form 
\begin{equation}\label{vis/den}
\frac{\eta}{s}=\frac{1}{4 \pi}\sqrt{\frac{1+\lambda_1G}{{\mathcal{Z}_1}}}=\frac{1}{4 \pi}\sqrt{\frac{1+\lambda_1 G(X)}{1+ \lambda_1 G(X) + \lambda_2X A_2(X)}}, 
\end{equation}
or 
\begin{equation}\label{vis/den2}
\frac{\eta}{s}=\frac{1}{4 \pi}\sqrt{\frac{{\mathcal{Z}_1(X) \Big{|}_{\lambda_2=0}}}{{\mathcal{Z}_1(X)}}}.
\end{equation}
{Here, we note that there is no a presence of the location of the event horizon $r_h$ on the $\eta/s$ ratio, and although the dimension of the space-time $D$ is not present in (\ref{vis/den}) (or (\ref{vis/den2})), this expression appears actively from the relation of the coupling functions (\ref{eq:mastereq}). Together with the above, since
\begin{equation}\label{eq:Z1_cond}
\mathcal{Z}_1(X):=1+ \lambda_1 G(X) + \lambda_2X A_2(X) >0,
\end{equation}
from the positivity of the mass $\mathcal{M}$ (and the entropy $\mathcal{S}_{W}$), and since 
 \begin{equation}\label{eq:G1_cond}
1+ \lambda_1 G(X) \ge 0, 
\end{equation}
from the reality and non-negativity of the shear viscosity $\eta$, we conclude that  
$\lambda_2XA_2(X)> -(1+ \lambda_1 G(X)) \le 0$.} {It is worth pointing out that the above conditions are not new, in fact, according to \cite{Takahashi:2019oxz}, the condition (\ref{eq:Z1_cond}) together with the strict inequality from (\ref{eq:G1_cond}) allow us to obtain necessary conditions of stability in four-dimensional spherical symmetric solutions. Curiously enough, the above is consistent with the stability conditions for linear cosmological perturbations, guaranteeing that both the effective gravitational constant and the squared propagation speed of the tensor modes are positive \cite{Langlois:2017mxy}, without the requirement on the sign of $\lambda_2X A_2(X)$. For our particular situation, and given the explicit expression for the scalar field $\phi$ found in (\ref{eq:phi}), where $X>0$, only the possibilities of study are performed by analyzing $\lambda_2 A_2(X)$. First, if $\lambda_2A_2(X)>0$, the KSS bound would be violated  even in the limit $\lambda_1 \to 0$. On the other hand, if  $\lambda_2A_2(X)<0$, and still obeying the condition (\ref{eq:Z1_cond}), one recovers the usual KSS bound where ${\eta}/{s}>{1}/{(4 \pi)}$. Finally, if  $\lambda_2A_2(X)=0$, which can be achieved in the limit $\lambda_2\to 0$, the KSS bound is saturated (${\eta}/{s}={1}/{(4 \pi)}$).}

{As concrete examples from the cases showed before, we can see that the Einstein-Hilbert case, together with a cosmological constant (this is for $\lambda_i=0$ for $i\in\{1,2,3,4,5\}$, and $Z=-2\Lambda/\lambda_0$), is naturally recovered, where we can see that $\eta/s=1/(4\pi)$. Together with the above, for example, for 
$$G(X)=X^{j},\qquad A_2(X)=G_{X}=j X^{j-1},$$
where $j$ is a positive constant, while that $Z$ and $\mathcal{Z}_{2}$ satisfy the condition (\ref{eq:mastereq}), we obtain
$$0<\frac{1+\lambda_1X^{j}}{1+(\lambda_1+j \lambda_2) X^{j}}<\frac{1}{4\pi},$$
as long as $\lambda_2>0$.
}

{In resume, with all this information we can to conclude that,  for a higher dimensional scalar-tensor theory (\ref{action})-(\ref{L45})  with specific coupling functions, represented via $Z$, $G$ and the $A_i's$, the explicit expression for the kinetic term $X$ can be obtained through (\ref{eq:mastereq}), and the solution takes the form (\ref{10})-(\ref{eq:L2}). Moreover, in particular, there exists an active presence of the functions $G$ and $\mathcal{Z}_1$ in the $\eta/s$ ratio, providing a new example of the violation of the KSS bound whose Lagrangian is at most linear in curvature tensor.
}

{\section{Conclusions and discussions}
\label{conclusions}}

In the present paper we explored the dimensional continuation of planar hairy black hole solutions found in \cite{Baake:2020tgk} and \cite{Santos:2020lmb}, where the theory is given by a model denominated as DHOST theory, Eqs.  (\ref{action})-(\ref{L45}), constructed by a non-trivial scalar field $\phi$ and its derivatives, the Ricci scalar $R$, and coupling functions depending on the kinetic term $X:=\partial_{\mu}\phi\,\partial^{\mu}\phi$ Eq.  (\ref{eq:X}), which we suppose to be constant. In this case, black holes resemble the well-known Schwarzschild Anti-de Sitter configurations in arbitrary dimensions, where the integration constant $M$ is related to the mass, and the AdS-radius depends on the coupling functions present in the theory, being interpreted as an effective cosmological constant. With these results, via the Wald formalism, we compute their thermodynamical parameters, which satisfy the First Law, Eq.  (\ref{eq:first-law}), as well as a higher dimensional Smarr relation, Eq.  (\ref{eq:smarr}).

Motivated by recent concrete examples presented in the literature (see \cite{Feng:2015oea,Brito:2019ose,Figueroa:2020tya,Bravo-Gaete:2021hlc,Santos:2020lmb}), where the bound for the viscosity/entropy density ratio Eq. (\ref{KSS}) can be violated, we analyzed the shear viscosity $\eta$ for DHOST theory, Eqs.  (\ref{action})-(\ref{L45}), following two procedures.  The first one, through the construction of a conserved charge as well as a suitable election of the Killing vector \cite{Fan:2018qnt} {where in this case is not necessary to impose any hydrodynamic limit, such as the low frequencies, to define the transport coefficient. On} the other hand, in the second one, via Green's functions and Kubo formula, given by \cite{Son:2002sd} and \cite{Brigante:2007nu} respectively, {and explained in the Appendix \ref{Kubo}.}  {For both techniques, we obtain the same expression for the $\eta/s$ ratio. Here we note that these results are not a surprise, because the boundary condition to the effective action for the transverse off-diagonal gravitons $h_{x_1 x_2}$ for the two methods considers the gravity fluctuations around the metric. For higher dimensional planar black holes in DHOST theories, the presence of the coupling functions $G$ and $\mathcal{Z}_1$ in the $\eta/s$ ratio, provides a new example of the violation of the KSS bound whose Lagrangian is at most linear in curvature tensor. Note that only with specific coupling functions (these are $Z$, $G$ and the $A_i's$),  the explicit expression for the kinetic term $X$, obtained through (\ref{eq:mastereq}), and the solution (\ref{10})-(\ref{eq:L2}), are required.} In fact,  according to the expression (\ref{vis/den}) (or (\ref{vis/den2})), the $\eta/s$ ratio is controlled by the parameters $\lambda_1$ and $\lambda_{2}$  as well as the kinetic term $X$ and the coupling functions $A_{2}(X)$ together with $G(X)$, allowing us to get cases where the KSS bound can be {violated, fulfilled or  saturated.} It is worth pointing out that although in the expression (\ref{vis/den}) there is no explicit presence of the dimension of the space-time, this quantity appears implicitly in the relation of the coupling functions obtained in (\ref{eq:mastereq}).

From this work, some natural extensions can be raised. For example, to simplify our computations, from the beginning we suppose that the kinetic term $X$ is a constant. It would be interesting to explore more general solutions in arbitrary dimensions, where now from (\ref{eq:X}) $X=X(r)$, and allowing us to explore the shear viscosity $\eta$ on configurations with non-standard asymptotically behaviors, as was studied for Lifshitz black holes in \cite{Brito:2019ose}. {Finally, these results deserve further investigation both in their own right, in particular in the context of the AdS/CFT correspondence and their implications.}

\bigskip 

\begin{acknowledgments}
{The authors would like to thank the Referee for the commentaries and suggestions to improve the paper. M.B. is supported by PROYECTO INTERNO UCM-IN-22204, L\'INEA REGULAR. 
H.B.-F. would like to thank partial financial support from Brazilian agencies  Coordenação de Aperfeiçoamento de Pessoal de Nível Superior (CAPES), under finance code 001, and Conselho Nacional de Desenvolvimento Científico e Tecnológico (CNPq) under Grant No. 311079/2019-9.   
}
\end{acknowledgments}

\appendix

\section{\large Relevant tensors and vectors for the equations of motion} \label{Appendix}
{For the sake of completeness, in this Appendix we report the expressions for ${\cal{G}}^{Z}_{\mu\nu}
,{\cal{G}}^{G}_{\mu\nu}$, the ${\cal{G}}^{(i)}_{\mu\nu}$'s and $\mathcal{J}^{\mu}$ present in the equations (\ref{eq:emotion1})-(\ref{eq:emotion2})}
\begin{align*}
{\cal{G}}^{Z}_{\mu\nu}&=\lambda_0 \left(-\frac{1}{2} Z(X) g_{\mu\nu}+Z_{X} \phi_{\mu}
\phi_{\nu}\right), 
\\ 
{\cal{G}}^{G}_{\mu\nu}&= (1+\lambda_1 G)  G_{\mu\nu}+\lambda_1 G_{X} R \phi_{\mu} \phi_{\nu} -\lambda_1 \nabla_{\nu} \nabla_{\mu} G  \\
&+\lambda_1 g_{\mu\nu} \nabla_{\lambda}\nabla^{\lambda}  G,
\\ 
{\cal{G}}^{(2)}_{\mu\nu}&=\lambda_2 \Big[-\phi_{\mu}\,(A_{2X} \nabla_{\nu}X)\,\Box \phi-(A_{2X} \nabla_{\mu} X ) \phi_{\nu} \,\Box \phi\\
&-A_2  \phi_{\nu \mu} \Box \phi -\phi_{\nu \mu} \phi_{\lambda} ( A_{2X} \nabla^{\lambda}
X)
\\
&+\phi_{\nu} \phi_{\lambda \mu} (A_{2X} \nabla^{\lambda}X)+\phi_{\mu}
\phi_{\lambda\nu} (A_{2X} \nabla^{\lambda} X)\\
&+A_2  R_{\nu\lambda} \phi_{\mu} \phi^{\lambda}+A_2  R_{\mu\lambda} \phi_{\nu}
\phi^{\lambda}\\
&-A_2  \phi_{\lambda\nu\mu} \phi^{\lambda}+\frac{1}{2}A_2  g_{\mu\nu} (\Box \phi)^2\\
&+g_{\mu\nu} \phi_{\lambda}  (A_{2X} \nabla^{\lambda} X) \Box \phi +A_2  g_{\mu\nu}
\phi^{\lambda}  \phi^{\,\,\rho \,}_{\rho \,\,\lambda}
\\
&-A_2  g_{\mu\nu} R_{\lambda\rho} \phi^{\lambda} \phi^{\rho}
+\frac{1}{2} A_2  g_{\mu\nu} \phi_{\rho\lambda} \phi^{\rho\lambda}\\
&+A_{2X} \phi_{\mu} \phi_{\nu}  \big((\Box \phi)^2-\phi_{\lambda\rho}
\phi^{\lambda\rho}\big)\Big],
\\
{\cal{G}}^{(3)}_{\mu\nu}&=\lambda_3 \Big[-\frac{1}{2}A_3  \phi_{\mu} \phi_{\nu} (\Box
\phi)^2-
\frac{1}{2} \phi_{\mu} \phi_{\nu} \phi_{\lambda}  (A_{3X} \nabla^{\lambda}X) \Box \phi\\
&+\frac{1}{2} A_{3}  \phi_{\mu}  \phi_{\lambda\nu} \phi^{\lambda} \Box \phi+
\frac{1}{2} A_{3}  \phi_{\nu} \phi_{\lambda\mu}\phi^{\lambda}
\Box \phi\\
&-\frac{1}{2} A_{3}  \phi_{\mu} \phi_{\nu} \phi^{\lambda} \phi_{\rho\,\, \lambda}^{\,\,
\rho \,} +\frac{1}{2} A_3  R_{\lambda\rho} \phi_{\mu} \phi_{\nu} \phi^{\lambda}
\phi^{\rho}\\
&-\frac{1}{2} \phi_{\mu}  (A_{3X} \nabla_{\nu} X) \phi^{\lambda}  \phi_{\rho\lambda}
\phi^{\rho} -\frac{1}{2} (A_{3X} \nabla_{\mu} X )\phi_{\nu}
\phi^{\lambda} \phi_{\rho\lambda} \phi^{\rho} \\
&-\frac{1}{2}A_3  \phi_{\nu} \phi^{\lambda} \phi_{\rho\lambda\mu}
\phi^{\rho}-\frac{1}{2}A_3
\phi_{\mu} \phi^{\lambda}  \phi_{\rho\lambda\nu} \phi^{\rho}\\
&-A_3  \phi_{\nu} \phi^{\lambda} \phi_{\rho\lambda} \phi^{\rho}_{\,\,\mu}-A_3  \phi_{\mu}
\phi^{\lambda}
\phi_{\rho\lambda} \phi^{\rho}_{\,\ \nu} \\
&+\frac{1}{2} g_{\mu\nu} \phi_{\lambda} (A_{3X} \nabla^{\lambda} X) \phi^{\rho}
\phi_{\sigma\rho}  \phi^{\sigma} +\frac{1}{2} g_{\mu\nu} A_{3}  \phi^{\lambda}  \phi^{\rho}
\phi_{\sigma\rho\lambda} \phi^{\sigma}\\
&+g_{\mu\nu} A_{3}  \phi^{\lambda}  \phi^{\rho} \phi_{\sigma\rho}
\phi^{\sigma}_{\,\,\lambda}+A_{3X} \phi_{\mu}
\phi_{\nu} (\Box \phi) \phi^{\rho} \phi_{\sigma\rho} \phi^{\sigma}\Big], 
\\
{\cal{G}}^{(4)}_{\mu\nu}&=\lambda_4\Big[-A_4  \phi_{\mu} \phi_{\nu} \phi^{\lambda}
\phi_{\rho\,\,\lambda}^{\,\,\rho}+ A_4  \phi_{\lambda\mu} \phi^{\lambda} \phi_{\rho\nu} \phi^{\rho}
\\
&-\phi_{\mu}  \phi_{\nu}  (A_{4X} \nabla^{\lambda} X) \phi_{\rho\lambda} \phi^{\rho} -A_4
\phi_{\mu} \phi_{\nu}
\phi_{\rho\lambda} \phi^{\rho\lambda}\\
&-\frac{1}{2} A_4  g_{\mu\nu} \phi^{\lambda} \phi^{\rho} \phi_{\sigma\rho}
\phi^{\sigma}_{\,\,\lambda} +A_{4X} \phi_{\mu} \phi_{\nu} \phi_{\lambda\rho} \phi^{\lambda}
\phi^{\rho\sigma} \phi_{\sigma}\Big],
\end{align*}

\begin{align*}
{\cal{G}}^{(5)}_{\mu\nu}&=\lambda_5\Big[-A_5  \phi_{\mu} \phi_{\nu} \phi^{\lambda} \phi_{\rho\lambda}
\phi^{\rho} (\Box \phi)
-\phi_{\mu} \phi_{\nu}  \phi_{\lambda} (A_{5X} \nabla^{\lambda} X) \phi^{\rho} \phi_{\sigma\rho} \phi^{\sigma}\\
&+A_5  \phi_{\nu}  \phi_{\lambda\mu} \phi^{\lambda} \phi^{\rho} \phi_{\sigma\rho} \phi^{\sigma}+
A_5  \phi_{\mu} \phi_{\lambda\nu} \phi^{\lambda} \phi^{\rho}
\phi_{\sigma\rho} \phi^{\sigma}\\
&-A_5  \phi_{\mu} \phi_{\nu}  \phi^{\lambda} \phi^{\rho}
\phi_{\sigma\rho\lambda} \phi^{\sigma}-2A_5  \phi_{\mu} \phi_{\nu}  \phi^{\lambda} \phi^{\rho} \phi_{\sigma\rho} \phi^{\sigma}_{\,\,\lambda}\\
&-\frac{1}{2} A_5  g_{\mu\nu} \phi^{\lambda} \phi_{\rho\lambda} \phi^{\rho} \phi^{\sigma}
\phi_{\tau\sigma} \phi^{\tau}+A_{5X} \phi_{\mu}  \phi_{\nu}
 \phi^{\lambda} \phi^{\rho} \phi_{\rho\lambda}
\phi^{\sigma}\phi^{\tau} \phi_{\tau\sigma}\Big],
\end{align*}
while that
\begin{eqnarray*}
\mathcal{J}^{\mu}=\mathcal{J}^{\mu}_{Z}+\mathcal{J}^{\mu}_{G}+\sum_{i=2}^{5} \mathcal{J}^{\mu}_{(i)},
\end{eqnarray*}
with
\begin{eqnarray*}
\mathcal{J}^{\mu}_{Z}&=&2 \lambda_0 Z_{X} \phi^{\mu}, \\ 
\mathcal{J}^{\mu}_{G}&=&2 \lambda_1 G_{X} R \phi^{\mu}, \\
\mathcal{J}^{\mu}_{(2)}&=& \lambda_2 \Big\{2 A_{2X} \phi^{\mu}\left[ (\Box \phi)^{2}-
\phi_{\lambda \rho}\phi^{\lambda \rho}\right]-2\nabla_{\nu} \left[ A_2  \left( g^{\mu \nu}-\phi^{\mu \nu}\right)\right]\Big\},
\\ 
\mathcal{J}^{\mu}_{(3)}&=&\lambda_3 \Big\{2 A_{3X} \phi^{\mu}  \, \Box \phi\,  \phi^{\lambda}
\phi_{\lambda \rho} \phi^{\rho}+2 A_3  \,\Box \phi \,
\phi^{\mu}_{\,\,\,\lambda}\phi^{\lambda}\\
&-&\nabla_{\nu} \left[A_3  \big(g^{\mu \nu} \phi^{\lambda} \phi_{\lambda \rho} \phi^{\rho}+\Box \phi \,\phi^{\mu} \phi^{\nu}\big)\right]\Big\},\\ 
\mathcal{J}^{\mu}_{(4)}&=&\lambda_4 \Big\{2 A_{4X} \phi^{\mu} \phi^{\sigma} \phi_{\sigma \rho}
\phi^{\rho \lambda}
\phi_{\lambda}+A_4(X) \big[\phi^{\mu}_{\,\,\rho} \phi^{\rho \lambda} \phi_{\lambda}\\
&+&\phi^{\sigma} \phi_{\sigma \rho} \phi^{\rho \mu} \big]-\nabla_{\nu}\big[A_4(X) \big(\phi^{\mu} \phi^{\nu \rho} \phi_{\rho}\\
&+&\phi^{\sigma} \phi_{\,\,\,\sigma}^{\mu}
\phi^{\nu}\big)\big]\Big\}, \\ 
\mathcal{J}^{\mu}_{(5)}&=&\lambda_5 \Big\{2 A_{5X} \phi^{\mu} \big( \phi^{\sigma} \phi_{\sigma
\rho} \phi^{\rho}\big)^2 +2 A_5(X)\big(
\phi^{\sigma} \phi_{\sigma \rho} \phi^{\rho}\big)\big(\phi^{\mu \sigma}\phi_{\sigma}\\
&+&\phi^{\sigma \mu} \phi_{\sigma}\big)-2\nabla_{\nu}\left[A_5(X)
\phi^{\sigma} \phi_{\sigma \rho} \phi^{\rho}  \phi^{\mu}
\phi^{\nu}\right]\Big\}.
\end{eqnarray*}

\bigskip

\section{\large Shear viscosity from Green's functions and Kubo formula} \label{Kubo} 

In order to corroborate the above computation, we perform {in this section} transverse and traceless perturbations following the steps of \cite{DeWolfe:1999cp,Csaki:2000fc,Gasperini:2007zz,Brito:2019ose,Santos:2020egn}. In the gravity side, we have that the black hole in generalized scalar-tensor theories in arbitrary dimensions plays the role of the gravitational dual of a certain fluid. Besides, to compute the shear viscosity through the holographic correspondence {it} is necessary {to}  linearize the field equations, as in \cite{Kovtun:2004de,Sadeghi:2018vrf,Kovtun:2003wp},  so that the effective hydrodynamics in the boundary field theory can be constructed using conserved currents and the energy-momentum tensor. In this sense, we do not consider the scalar field perturbations, { {\sl i.e.}}  $\delta\phi=0$. Thus, the original Ricci tensor of the background metric acquires a single non-vanishing correction at linear order in $\Psi$, as  in Eq. \eqref{pert_metric}
\begin{equation}
R^{(1)}_{x_{1}x_{2}}=\frac{1+\lambda_1 G}{\mathcal{Z}_1}\left(-\frac{r^{2}}{2}\Box\Psi-rf^{'}\Psi-(D-2)f\Psi\right),\label{T}
\end{equation}
where we can identify that 
\begin{equation}
R^{(0)}_{xx}=-rf^{'}-(D-2)f.\label{T1}
\end{equation}
Here, the equation (\ref{T1}) denotes any of the (diagonal) components of the zeroth-order Ricci tensor of the  background metric. {Combining equations (\ref{T}) and (\ref{T1}), we have
\begin{equation}
R^{(1)}_{x_{1}x_{2}}=\frac{1+\lambda_1G}{\mathcal{Z}_1}\left(-\frac{r^{2}}{2}\Box\Psi+R^{(0)}_{xx}\Psi\right),\label{T2}
\end{equation}}
and we can write the {perturbed} Einstein tensor {to first order} as:
\begin{eqnarray}
G^{(1)}_{x_{1}x_{2}}&=&\frac{1+\lambda_1G}{\mathcal{Z}_1}\left(-\frac{r^{2}}{2}\Box\Psi+R^{(0)}_{xx}\Psi-\frac{1}{2}R^{(0)}\Psi\right)\\
                    &=&\frac{1+\lambda_1G}{\mathcal{Z}_1}\left(-\frac{r^{2}}{2}\Box\Psi+G^{(0)}_{xx}\Psi\right).\label{T3}
\end{eqnarray}
Together with the above {and considering the Einstein tensor as given by $G^{(0)}_{xx}=T^{(0)}_{xx}/2$}, we can write the equation (\ref{T3}) as  
\begin{eqnarray}
&&\frac{1+\lambda_1G}{\mathcal{Z}_1}\left(\Box\Psi-\frac{1}{r^2}(T^{(0)}_{xx}+T^{(1)}_{x_{1}x_{2}})\right)\nonumber\\
&&=\sqrt{\frac{1+\lambda_1G}{\mathcal{Z}_1}}\left(\sqrt{\frac{1+\lambda_1G}{\mathcal{Z}_1}}\Box\Psi-\frac{1}{r^2}\sqrt{\frac{1+\lambda_1G}{\mathcal{Z}_1}}(T^{(0)}_{xx}+T^{(1)}_{x_{1}x_{2}})\right)=0,\label{T5}
\end{eqnarray}
where $T^{(1)}_{x_{1}x_{2}}=(\delta T_{x_{1}x_{2}}/\delta g_{x_{1}x_{2}})\delta g_{x_{1}x_{2}}$, with
\begin{eqnarray}
T^{(0)}_{\mu\nu}=-\frac{2}{\mathcal{Z}_1}\frac{\delta}{\delta g^{\mu\nu}}\left(\sum_{i=2}^{5} \lambda_i A_{i} {\mathcal{L}}_{i}\right)+\frac{g_{\mu\nu}}{\mathcal{Z}_1}\sum_{i=2}^{5} \lambda_iA_{i} {\mathcal{L}}_{i}.\label{T6}
\end{eqnarray}
Now, {for} the component $T^{(0)}_{xx}$ we have
\begin{eqnarray}
T^{(0)}_{xx}=-\frac{2}{\mathcal{Z}_1}\frac{\delta}{\delta g^{xx}}\left(\sum_{i=2}^{5} \lambda_i A_{i} {\mathcal{L}}_{i}\right)+\frac{g_{xx}}{\mathcal{Z}_1}\sum_{i=2}^{5} \lambda_i A_{i} {\mathcal{L}}_{i},\label{T7}
\end{eqnarray}
and using the Lagrangians
{\eqref{L23} and \eqref{L45} } 
we can see that all {the contributions from } ${\mathcal{L}}_{2}$, ${\mathcal{L}}_{3}$, ${\mathcal{L}}_{4}$ and ${\mathcal{L}}_{5}$ becomes null, because the kinetic coupling is a function of the radial component, namely, $\phi=\phi(r)$. In this case, $T^{(0)}_{xx}=0=T^{(1)}_{x_{1}x_{2}}$ and {we can write 
\begin{eqnarray}
\mathcal{Z}_1\sqrt{\frac{1+\lambda_1 G}{\mathcal{Z}_1}}\; \Box\, \Psi=0,\label{T8}
\end{eqnarray}
where in the metric background (\ref{7}) {with $h=f$}, and considering the equation (\ref{T8}), we have
\begin{eqnarray}
\mathcal{Z}_1\sqrt{\frac{1+\lambda_1 G}{\mathcal{Z}_1}}f\Psi^{''}+\mathcal{Z}_1\sqrt{\frac{1+\lambda_1 G}{\mathcal{Z}_1}}\left(f^{'}+\frac{(D-2)f}{r}\right)\Psi^{'}+\mathcal{Z}_1\sqrt{\frac{1+\lambda_1 G}{\mathcal{Z}_1}}\frac{\ddot{\Psi}}{f}=0.\label{T9}
\end{eqnarray}
Now, we consider the following {\sl ansatz}
\begin{eqnarray}
\Psi=\int{\frac{d^{(D-1)}k}{(2\pi)^{(D-1)}}}e^{ikx}\chi(r,k).\label{T10}
\end{eqnarray}
Here $x=(t,\vec{x}\,)$ and $k=(\omega,\vec{q}\,)$, where as we know in general for the mass term the equation (\ref{T10}) contains  contributions $k^2=q^2-\omega^2$. Nevertheless, {considering the case $\omega\to0$ and spatial momentum $q=0$, we can find that $\chi(r,k)=\chi(r)$}. In addition, we can write the equation (\ref{T10}) in terms of a Klein-Gordon-like equation \cite{Brito:2019ose,Santos:2020egn} as follows
\begin{eqnarray}
\frac{1}{\sqrt{-g}}\partial_{\alpha}\left(\mathcal{Z}_1\sqrt{\frac{1+\lambda_1 G}{\mathcal{Z}_1}}\sqrt{-g}g^{\alpha\beta}\partial_{\beta}\Psi\right)=0,\label{T11}
\end{eqnarray}
and effective action for the equation (\ref{T11}) can be written as 
\begin{eqnarray}
S=-\int{d^{(D-1)}k\left(\frac{N(r)}{2}\frac{d\chi(r,k)}{dr}\frac{d\chi(r,-k)}{dr}\right)}
\end{eqnarray}
where $N(r)=\mathcal{Z}_1\sqrt{(1+\lambda_1 G)/\mathcal{Z}_1}\sqrt{-g}g^{rr}$.} This action on-shell reduces to the surface term
\begin{eqnarray}\label{S_cl}
S=-\left.\int{d^{(D-1)}k\left(\frac12{N(r)}\chi(r,k){\partial_r\chi(r,-k)}\right)}\right|_{r_h}^{\infty}.
\end{eqnarray}
Following the procedure of \cite{Brito:2019ose,Santos:2020egn}, we can extract the retarded Green’s function, which reads
{\begin{eqnarray}\label{G_Rfinal}
G^{R}_{x_{1}x_{2},x_{1}x_{2}}(\omega,0)=-2\left(\mathcal{Z}_1\sqrt{\frac{1+\lambda_1 G}{\mathcal{Z}_1}}\sqrt{-g}g^{rr}\Big{|}_{r_{h}}\right)\chi(r,-\omega)\partial_{r}\chi(r,\omega)\Big{|}_{r_{h}}. 
\end{eqnarray}
{This expression for the Green's function diverges at the horizon. In order to remove this  divergence we implement a regularity condition such that the derivative of $\chi$  is given in terms of $\chi(r_h)$ at leading order in $\omega\to 0$. This low frequency limit corresponds to the hydrodynamical limit and has physical importance to define transport coefficients, such as the shear viscosity, in our case}  \cite{Chakrabarti:2010xy}. From equation (\ref{G_Rfinal}), we have 
\begin{eqnarray}
G^{R}_{x_{1}x_{2},x_{1}x_{2}}(\omega,0)=-{2i\omega r^{D-2}_{h}}\mathcal{Z}_1\sqrt{\frac{1+\lambda_1 G}{\mathcal{Z}_1}}=-2i\omega \left(\frac{1}{4 \pi} \sqrt{\frac{1+\lambda_1 G}{\mathcal{Z}_1}}\right) s, 
\end{eqnarray}
where the entropy density $s$ was given in  (\ref{s}). Finally, the shear viscosity $\eta$ \cite{Policastro:2001yc,Son:2002sd} is  given by
\begin{eqnarray}
\eta&=&-\lim_{\omega\rightarrow 0}\frac{1}{2\omega}{\rm Im}G^{R}_{x_{1}x_{2},x_{1}x_{2}}=\left(\frac{1}{4 \pi} \sqrt{\frac{1+\lambda_1 G}{\mathcal{Z}_1}}\right) s \nonumber\\
&\Rightarrow& \frac{\eta}{s}=\left(\frac{1}{4 \pi} \sqrt{\frac{1+\lambda_1 G}{\mathcal{Z}_1}}\right),\label{VS}
\end{eqnarray}
recovering the viscosity/entropy density ratio obtained in Eq.  (\ref{vis/den}).} According to \cite{Kovtun:2004de}, the right-hand side
of Eq. (\ref{VS}) is related to the absorption cross-section of low-energy gravitons. From usual theories that consider the low-frequency regime, the relation $\eta/s$ is not violated but has a universal bound. The main idea behind the computations of the viscosity/entropy density ratio is to characterize how close a given fluid is to be perfect. However, in our study we show that for the  generalized scalar-tensor theories in arbitrary dimensions, Eqs.  (\ref{action})-(\ref{L45}), this limit is violated, due to the coupling constants that control the influence of the kinetic term in this relation.


\end{document}